\documentclass[12pt]{iopart}
\usepackage[utf8]{inputenc}
\usepackage[T1]{fontenc}
\usepackage[ngerman,english]{babel}
\expandafter\let\csname equation*\endcsname\relax
\expandafter\let\csname endequation*\endcsname\relax
\usepackage{amsmath}
\usepackage{iopams}
\usepackage{graphicx}
\usepackage{braket}
\usepackage{float}
\usepackage{amssymb}
\usepackage{amsopn}
\usepackage{setspace}
\usepackage{epstopdf}
\usepackage{units}
\usepackage{hyperref}
\usepackage{lmodern}
\usepackage{xspace}

\usepackage[numbers, sort&compress]{natbib}
\usepackage{doi}
\usepackage{sidecap}
\hypersetup{colorlinks=true, linkcolor=blue, citecolor=blue, urlcolor=blue}
\def\newblock{\hskip .11em plus .33em minus .07em} 

\begin{document}
\title[High-order above-threshold photoemission from nanotips in two-color laser fields]{High-order above-threshold photoemission from nanotips controlled with two-color laser fields}
\author{Lennart Seiffert$^{1}$, Timo Paschen$^{2}$, Peter Hommelhoff$^{2}$ and Thomas Fennel$^{1,3,*}$}
\address{$^1$Institute of Physics, University of Rostock, Albert-Einstein-Straße 23, D-18059 Rostock, Germany}
\address{$^2$Department of Physics, Friedrich-Alexander-Universit\"at Erlangen-N\"urnberg (FAU), Staudtstra{\ss}e 1, 91058 Erlangen, Germany}
\address{$^3$Max-Born-Institut, Max-Born-Straße 2A, D-12489 Berlin, Germany}
\eads{\mailto{\color{blue}thomas.fennel@uni-rostock.de}}

\begin{abstract}
We investigate the process of phase-controlled high-order above-threshold photoemission from sharp metallic nanotips under bichromatic laser fields. Experimental photoelectron spectra resulting from two-color excitation with a moderately intense near-infrared fundamental field (1560\,nm) and its weak second harmonic show a strong sensitivity on the relative phase and clear indications for a plateau-like structure that is attributed to elastic backscattering. To explore the relevant control mechanisms, characteristic features, and particular signatures from near-field inhomogeneity, we performed systematic quantum simulations employing a one-dimensional nanotip model. Besides rich phase-dependent structures in the simulated above-threshold ionization (ATI) photoelectron spectra we find ponderomotive shifts as well as substantial modifications of the rescattering cutoff as function of the decay length of the near-field. To explore the quantum or classical nature of the observed features and to discriminate the two-color effects stemming from electron propagation and from the ionization rate we compare the quantum results to classical trajectory simulations. We show that signatures from direct electrons as well as the modulations in the plateau region mainly stem from control of the ionization probability, while the modulation in the cutoff region can only be explained by the impact of the two-color field on the electron trajectory. Despite the complexity of the phase-dependent features that render two-color strong-field photoemission from nanotips intriguing for sub-cycle strong-field control, our findings support that the recollision features in the cutoff region provide a robust and reliable method to calibrate the relative two-color phase.
\end{abstract}

\noindent{\it Keywords\/ above-threshold ionization, two-color laser fields, nanotips, electron recollision, near-fields, phase control}
\maketitle
\section{Introduction}
When a point-like system, such as an atom or a molecule, is exposed to an intense laser field, electrons released via tunneling are accelerated by the field and can either be emitted directly or revisit the parent ion to recombine or rescatter~\cite{Corkum_PRL71_1993, Schafer_PRL70_1993, Krausz_RMP81_2009}. This famous and intuitive three-step picture turned out to be a central and stunningly robust concept in strong-field physics and has become a key paradigm of attosecond science~\cite{Corkum_NP3_2007}. While the recombination step gives rise to high harmonic generation~\cite{Lewenstein_PRA49_1994} and the formation of attosecond pulses~\cite{Sansone_Science314_2006, Goulielmakis_Science320_2008}, rescattering is responsible for the high energy electron emission. In atomic strong-field ionization, the most-energetic electrons emerge from elastic backscattering with the well-known classical energy cutoff at $10\,U_\text{p}$, where $U_\text{p}$ ist the ponderomotive potential~\cite{paulus_PRL72_1994}.\\

In the strong-field ionization of atoms and molecules, collective effects are usually negligible and the driving laser field can be assumed to be homogenous on the spatial scale of the electron excursion such that the Coulomb field of the residual ion and electron correlations determine the fine structure of photoelectron spectra~\cite{Blaga_NP5_2009, Yan_PRL105_2010, Liu_PRL105_2010, Huismans_Science331_2011}. In the case of laser-driven nanostructures, the situation is different, as the optical near-field can be strongly enhanced and field inhomogeneities on the sub-wavelength scale can unfold substantial impact on the electron dynamics. A key demonstration is the quenching of the field-driven quiver motion using extremely localized near-fields at nanotips~\cite{Herink_Nature483_2012}. However, if the near-field extension is sufficiently large, laser-driven electron backscattering was not only found to generate the most energetic electrons for nanostructures as well, but also turned out to be well controllable with the carrier-envelope phase of the incident few-cycle laser pulse~\cite{Zherebtsov_NatPhys7_2011, Krueger_Nature475_2011, Piglosiewicz_NatPhot8_2014}. The presence and dominance of elastic backscattering documents the important conceptual link between strong-field atomic physics and strong-field nanoscience~\cite{Hommelhoff_Wiley_2015, Ciappina_RPP80_2017,Schoetz_PRA97_2018} and motivates the analysis of its significance for sub-cycle nanotip photoemission and resulting protocols for attosecond electronics based on waveform-controlled laser fields~\cite{Krausz_RMP81_2009}.\\

Recently, control of strong-field processes with bichromatic laser fields has become particularly popular~\cite{Ray_PRA83_2011, Skruszewicz_PRL115_2015, Eicke_JMO64_2017, Kerbstadt_NJP19_2017}, as two-color fields containing a strong fundamental and a weaker higher harmonic are easy to generate from frequency up-conversion. In particular, the intrinsic stability and controllability of the relative phase makes two-color fields attractive for strong-field waveform control~\cite{Shafir_NP5_2009}. Despite a large set of studies on atoms and molecules, two-color strong-field ionization of nanostructures has been studied only little so far. Recently, the resonant two-color excitation of pre-expanded clusters has been reported to enable sub-cycle controlled electron acceleration via plasmon assisted forward rescattering~\cite{Passig_NC8_2017}. Other studies targeted the non-destructive regime and reported strong phase dependence of the two-color ionization of tungsten nanotips~\cite{Foerster_PRL117_2016, Paschen_JMO64_2017}. The particularly high relevance of nanotip-based strong-field effects for applications motivates a systematic analysis of the relevant control mechanisms under bichromatic fields and the investigation of their sensitivity to near-field localization.\\

\begin{figure}[t!]
\centering
\includegraphics[width=0.5\textwidth]{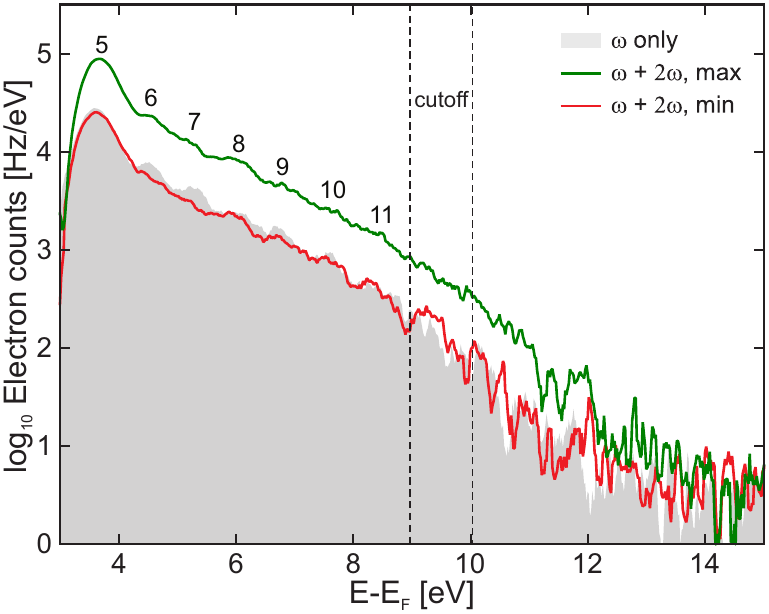}
\caption{Measured electron energy spectra from a nanotip in a bichromatic laser field with peak intensities $I_{\omega} = \unit[1.2 \times 10^{12}]{Wcm^{-2}}$ and $I_{2\omega} = \unit[3.1 \times 10^{10}]{Wcm^{-2}}$ (including field enhancement). Gray area: Only fundamental pulses focused onto the nanotip. Red: Fundamental and second harmonic with relative phase locked at the photocurrent minimum. Green: Fundamental and second harmonic with relative phase locked at the photocurrent maximum. Numbers indicate multi-photon orders of the fundamental, the dashed lines visualized the high-energy cutoff region. Energies are given with respect to the Fermi level $E_{\mathrm F}$.}
\label{fig1}
\end{figure}

 Our current theoretical analysis is motivated by a particular experimental two-color nanotip photoemission scenario similar to previous work~\cite{Foerster_PRL117_2016, Paschen_JMO64_2017}. A sharp tungsten nanotip with apex radius of $\unit[10]{nm}$ is illuminated with femtosecond pulses emitted from an Erbium-doped fiber laser. The fundamental has a central wavelength of $\unit[1560]{nm}$ and $\unit[74]{fs}$ pulse duration. The second harmonic field with wavelength $\unit[780]{nm}$ has a slightly shorter pulse duration of $\unit[64]{fs}$. The relative phase between the two colors is interferometrically stabilized and can be locked to an arbitrary relative phase value. A selection of measured photoelectron spectra is displayed in figure~\ref{fig1}. All spectra show a strong low energy peak followed by an exponentially decaying contribution that exhibits an increased slope beyond $\unit[10]{eV}$. The different spectra correspond to: Only the fundamental (gray area), both colors in phase-locked mode at the photocurrent minimum (red), and the phase-locked case at the photocurrent maximum (green). The respective near-field peak intensities include field enhancement factors of $\gamma_{\omega} = 7$ and $\gamma_{2\omega} = 6$. The nanotip used in the experiment is a monocrystalline tungsten tip oriented in [310] direction with work function $W_{310}=\unit[4.31]{eV}$~\cite{Mueller_JAP26_1955}. Here, electron energies are measured with respect to an effective barrier height of $W_\text{eff}=\unit[3.6]{eV}$, including the shift induced by the Schottky effect due to the bias voltage applied to the nanotip. Peaks indicated by numbers corresponding to the respective multi-photon order are clearly visible and the total photocurrent is homogeneously suppressed or enhanced in the case of the relative $\omega-2\omega$ phase locked to minimum or maximum photoemission, respectively. Additionally, the spectrum corresponding to phase-locked maximum exhibits a rather clearly visible high-energy cutoff (region indicated by the dashed lines). As will be motivated in more detail below, the stronger near-threshold peak and the plateau-like features are attributed to direct and backscattered electrons, respectively. In the presence of the weak $2\omega$ component, the suppression or strong enhancement of the signal when varying the relative phase documents a substantial two-color effect. The main goal of the following analysis is to explore the signatures and mechanisms of the phase-controlled nanotip photoemission in detail theoretically. A particular aspect is the analysis of the significance of near-field inhomogeneity effects and their relevance for future experiments.\\

In order to explore and analyze the photoemission dynamics, we considered a simplified one-dimensional metallic nanotip model and performed systematic sets of simulations that compare predictions resulting from quantum and classical treatments of the electron dynamics. In the quantum version, we describe the two-color photoemission by solving the one-electron time-dependent Schrödinger equation in the inhomogeneous near-field. Our model allows adjusting the essential nanotip parameters such as work function, Fermi energy, field enhancement and decay length. We find clear signatures for near-field induced ponderomotive shifts and pronounced two-color effects in the ATI spectra including a decay length-dependent cutoff. By the comparison with classical trajectory calculations we show that the phase-dependent signals from direct electrons and the signatures in the plateau region can be associated with ionization rate effects. The modulation of the high energy cutoff, however, can only be described by the two-color effect on the elastic backscattering trajectories. Our findings support that the phase-dependent recollision features provide a robust and useful marker for relative phase calibration.

\section{Methods}
\begin{figure}[h]
\centering
\includegraphics[width=0.8\textwidth]{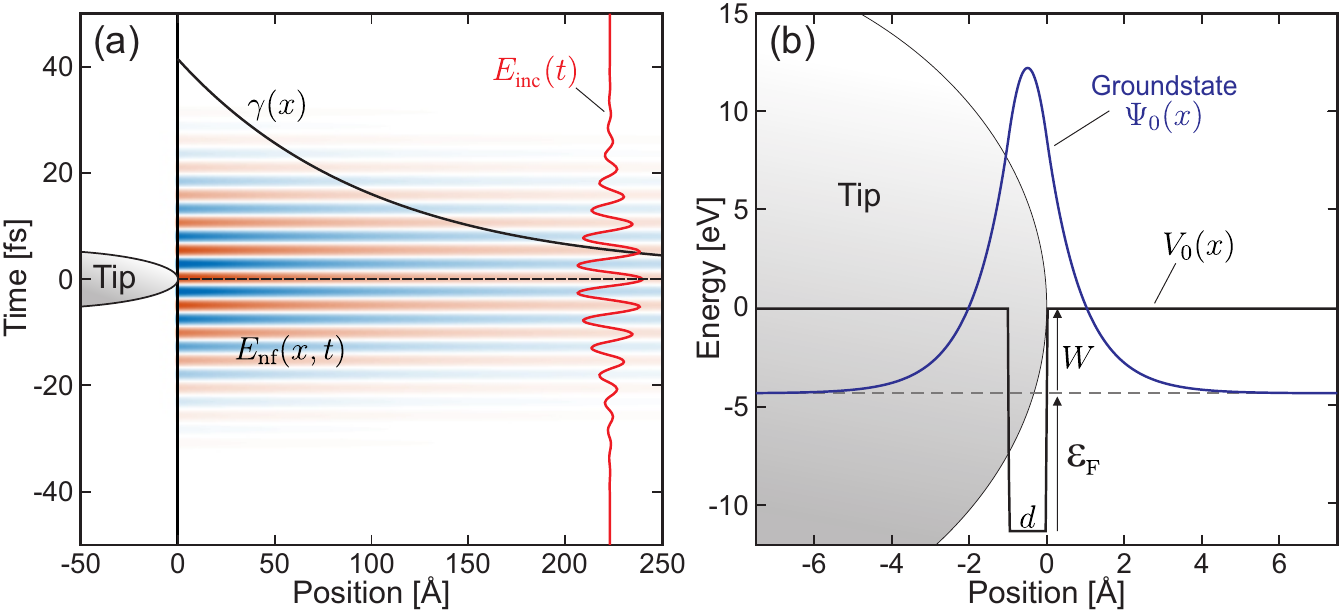}
\caption{(a) Schematics of the one-dimensional nanotip model with apex at $x=0$ and exponentially decaying field-enhancement $\gamma(x)$ (black curve) in the region $x\geq 0$. The evolution of the near-field $E_{\mathrm{nf}}$ (false color plot) results from assuming a local amplification of the incident laser-field $E_\text{inc}$ (red curve) via the enhancement profile. (b) Square-well potential inside the tip for the active electron, whose groundstate energy $\varepsilon_0$ (dashed line) matches the Fermi energy and work function of the material. }
\label{fig2}
\end{figure}

In the following we motivate the assumptions underlying the simplified one-dimensional nanotip model and discuss both its quantum mechanical implementation employing the solution of the time-dependent Schrödinger equation (TDSE) as well as the classical trajectory version~\citep{Krause_PRL68_1992, Corkum_PRL71_1993}. Common to both versions is the description of the effective field on the axis of the ideal metallic tip with apex at $x=0$. Inside the tip ($x<0$), the field is assumed to vanish because of perfect screening. On the surface and in the region outside the tip $(x\ge0)$, we consider a local instantaneous field enhancement of the incident laser field $E_\text{inc}(t)$ with an exponentially decaying profile. The resulting enhancement profile is defined as
\begin{equation}
\gamma(x)=\left\{\begin{array}{c c c} 0 &  \mbox{for} & x < 0  \\  1+ (\gamma_0-1)e^{-x/\lambda_\text{nf}} &\mbox{for}  & x \ge 0 \end{array}\right.
\end{equation}
with peak enhancement $\gamma_0$ and decay length $\lambda_\text{nf}$, that we assume to be equal for both spectral components. The bichromatic incident laser field is described as
\begin{equation}
E_\text{inc}(t)=E_0 f(t) \left[ \cos (\omega t)+\sqrt{\beta}\cos (2\omega t+\varphi)\right]
\end{equation}
with peak electric field amplitude of the fundamental $E_0$, normalized common Gaussian pulse envelope  $f(t)$, relative intensity of the $2\omega$-component $\beta$, and relative two-color phase $\varphi$. The resulting effective local near-field (see figure~\ref{fig2}(a)) reads
\begin{equation}
E_\text{nf}(x,t) = \gamma(x)E_\text{inc}(t)
\end{equation}
and is used for both model versions.

\subsection{Quantum mechanical description}
Starting point for the quantum mechanical implementation is the field-free ground state in a square well potential
\begin{equation}
V_0(x)=\left\{\begin{array}{c c c} \tilde V_0 &  \mbox{for} & -d < x < 0  \\  0 & & \text{otherwise} \end{array}\right.
\end{equation}
as illustrated in figure~\ref{fig2}(b). The width $d$ and depth ${\tilde V}_0$ are chosen such that two conditions are met. First, the ground state energy $\varepsilon_0$ must match $\varepsilon_0=\tilde V_0+\varepsilon_{\mathrm F}$ in order to reflect the Fermi-level of the considered material with Fermi energy $\varepsilon_{\mathrm F}$. Second, the binding energy must reflect the work function $\varepsilon_0=-W$. The ground state wave function $\Psi_0(x)$ is determined via imaginary time propagation. In the current study we use the work function $W=\unit[4.3]{eV}$~\cite{Mueller_JAP26_1955} and Fermi energy $\varepsilon_\text{F}=\unit[7.0]{eV}$~\cite{Mattheiss_PR139_1965} of bulk tungsten in [310]-direction. The resulting parameters are $d=0.99\AA$ and $\tilde V_0=-11.3\,\mathrm{eV}$. Besides matching the material parameters, this construction of the initial state supports no states that are more deeply bound, which mimics the Pauli-blocking of transitions to fully occupied states below the Fermi edge in a real metal. \\

The laser-driven dynamics is simulated by integrating the wavefunction $\Psi(x,t)$ according to the time-dependent Schrödinger equation
\begin{equation}
i\hbar\frac{\partial}{\partial t} \Psi(x,t) = \left[-\frac{\hbar^2}{2m}\frac{\partial^2}{\partial x^2} + V(x,t)\right] \Psi(x,t)
\end{equation}
using a Crank-Nicholson propagator. Here, $m$ is the electron mass, $\hbar$ is Planck's constant, and $V(x,t)$ is the effective time-dependent potential including the laser induced near-field in length gauge via
\begin{equation}
 V(x,t) = V_0(x) + e\int_0^x E_\text{nf}(x',t) dx'= V_0(x) + eE_\text{inc}(t)\int_0^x \gamma(x') dx'.
\end{equation}
Absorbing boundary conditions are implemented at both sides at the end of the numerical arena to prevent spurious reflections. The numerical arena itself is chosen large enough to accommodate all relevant spectral components for a sufficiently long time after the pulse to ensure converged electron spectra.  At the end of the simulation, energy spectra of electrons emitted to the right (i.e. into vacuum) are evaluated using the window operator method.

\subsection{Classical trajectory model}
For the purely classical description, trajectories are born at rest at the classical tunneling exit $x_\text{birth} = - W/E_\text{nf}(x=0,t_\text{birth})$. Trajectories are propagated in the two-color near-field via numerical integration of the classical equation of motion
\begin{equation}
\ddot{x}(t) = -\frac{e}{m}E_\text{nf}(x,t).
\end{equation}
For electrons returning to the surface ($x=0$), we consider elastic backscattering ($\dot{x} \rightarrow - \dot{x}$). Trajectories are weighted with an instantaneous tunneling rate~\cite{Ammosov_JETP_1986}. Energy spectra are evaluated from all emitted electrons, i.e. from electrons with ($x>0$) and positive final momenta. For historical reasons, this classical version is henceforward termed the simple man's model (SMM).

\section{Results and discussion}
In the following we assumed an incident fundamental $\omega$-field with wavelength $\unit[1560]{nm}$, pulse duration $\tau = \unit[20]{fs}$, intensity $I=\unit[1\times10^{11}]{Wcm^{-2}}$, a relative intensity of the $2\omega$-field of $\beta=0.01$, and a peak field enhancement of $\gamma_0 = 7$. As a result, the effective field intensity at the tip apex is close to $\unit[5\times10^{12}]{Wcm^{-2}}$. The resulting ponderomotive potential of the dominant $\omega$-component is $U_\text{p}^\text{inc} = \unit[0.02]{eV}$ in the incident field (without field enhancement) and $U_\text{p}^\text{apex} = \unit[1.14]{eV}$ at the tip. The corresponding electron excursion length in the fully enhanced near-field is $x_q = \unit[7.4]{\AA}$.\\

\begin{figure}[h]
\centering
\includegraphics[width=0.5\textwidth]{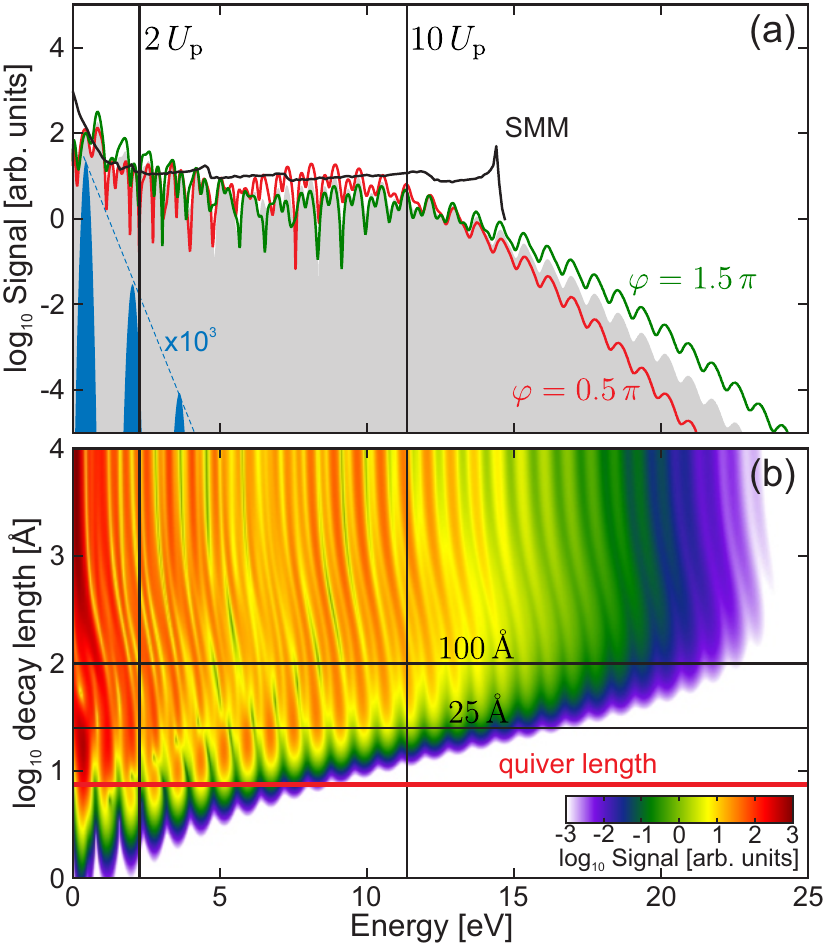}
\caption{(a) Energy spectra extracted from TDSE calculations including only the fundamental (gray area), only the second harmonic (blue area) and when using two-color fields with different relative phases (red and green). The solid black curve reflects the result of a respective SMM calculation for only the fundamental. Vertical black lines indicate the characteristic cutoff energies of direct and recollision electrons at $2$ and $10\,U_\text{p}$, respectively. (b) Map of ATI spectra  as function of the decay length of the near-field from the fundamental field only.}
\label{fig:tdse_spec}
\end{figure}

Figure~\ref{fig:tdse_spec} displays the individual and combined effects of the $\omega$ and $2\omega$ fields on the ATI spectra and illustrates the influence of the range of the near-field. The results in figure~\ref{fig:tdse_spec}(a) show that the $2\omega$ field alone induces only weak ATI peaks with the exponential intensity decrease typical for the perturbative regime (blue spectrum) and negligible net yield when compared to the $\omega$-only case (gray area). The latter exhibits the typical strong-field signatures such as a prominent direct electron feature at low energy followed by a plateau with a pronounced cutoff. In addition, the corresponding result from the classical SMM model is shown, which predicts a much sharper cutoff. In both cases, the plateau ends near 10\,$U_\text{p}$. Although the $2\omega$ pulse is very weak, its presence in the combined field results in strong, phase-sensitive modifications with respect to the $\omega$-only reference spectrum up to high energies (green and red lines). In particular, the enhancement shows a spectral chirp, i.e., for a given relative phase the emission current is amplified in the plateau region but suppressed in the cutoff region or vice versa. The detailed phase-dependent evolution will be discussed in more detail below.\\

To illustrate the strong influence of the range of the near-field, figure~\ref{fig:tdse_spec}(b) displays the evolution of ATI spectra for $\omega$-only excitation with decay length. The analysis reveals two main trends. On the one hand, the ATI spectrum quickly collapses in the limit of very strong field localization for decay lengths approaching the electron quiver amplitude (red horizontal line). In this limit, recollision becomes suppressed and eventually quenched~\cite{Schoetz_PRA97_2018}. The spectral positions of the ATI peaks, however, remain unchanged. For larger decay lengths ($\gtrsim\unit[100]{\AA}$), on the other hand, the ATI spectrum remains robust but ATI orders exhibits a gradual shift to lower energy with increasing decay length. This energy downshift has a magnitude of $U_\text{p}^\text{apex}$ and reflects the fact that the ponderomotive acceleration of electrons escaping from the near-field gradually vanishes. For intermediate values of the decay length ($\approx\unit[100]{\AA}$), the electrons can fully experience the inhomogeneous spatial profile of the near-field and can accumulate the corresponding ponderomotive energy during their escape before the temporal field envelope ceases. This is not possible for very long decay lengths, where the temporal profile will terminate the field long before the near-field decays, eliminating ponderomotive acceleration~\cite{Bucksbaum_JOSAB4_1987}. The underlying competition of the spatial and temporal decay effect also explains why low ATI orders show the shift already at much lower values of the near-field decay length than fast electrons. The latter can benefit much longer from the ponderomotive gain as they can still escape the region of the enhanced field during the pulse. This spectral dependence of the quenching of ponderomotive acceleration in the ATI spectrum represents a characteristic fingerprint of the inhomogeneity effect and could therefore serve as a sensitive marker for its experimental verification. For practical reasons, however, the variation of the pulse duration would be more attractive than changing the decay length in experiment.

\begin{figure}[h]
\centering
\includegraphics[width=1.0\textwidth]{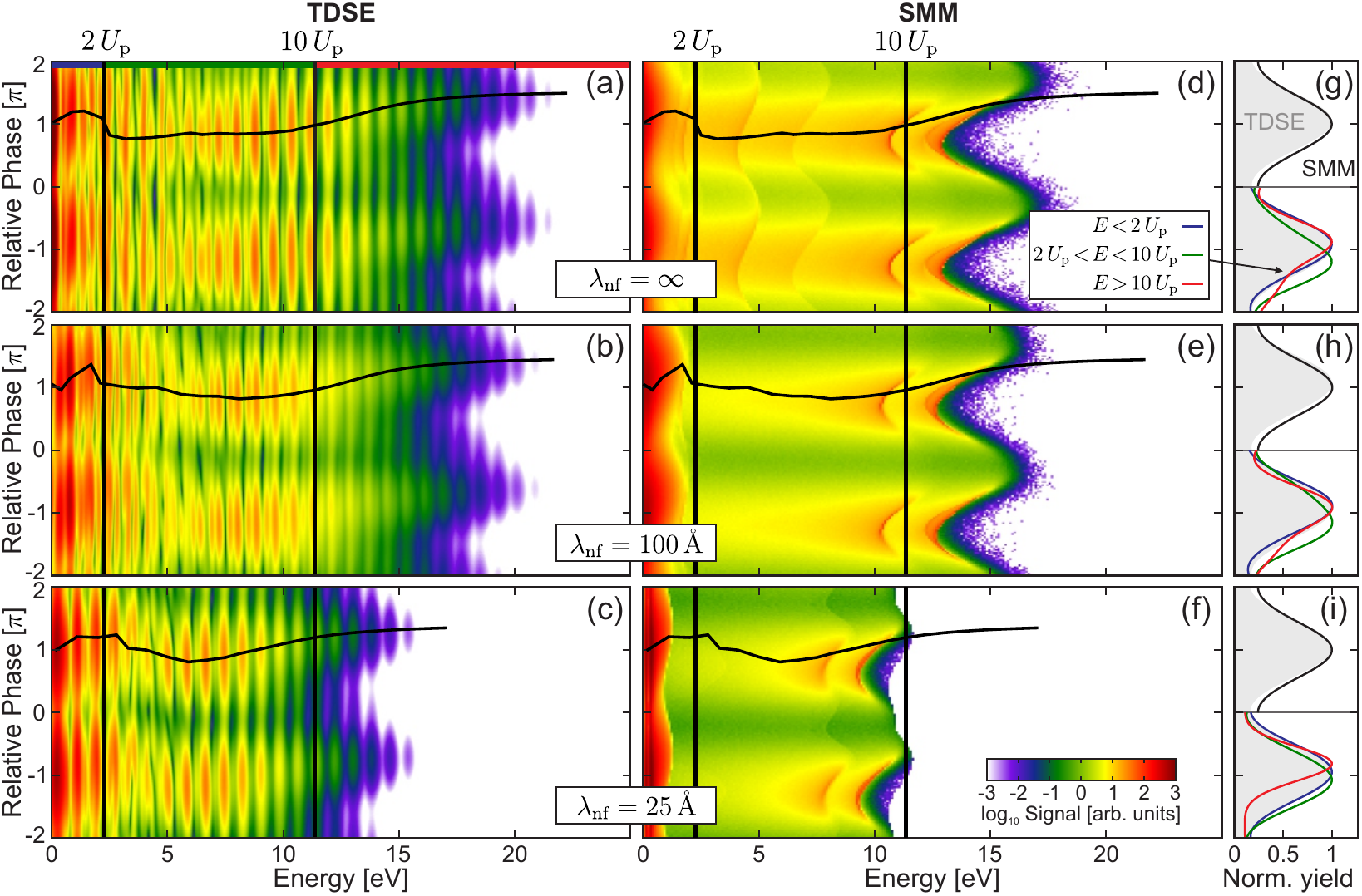}
\caption{Photoelectron spectra for two-color excitation as predicted from TDSE (a-c) and SMM (d-f) as function of relative phase for selected values of the decay length. Black curves indicate the evolution of the critical phases extracted from the TDSE results. Note that for comparison these curve are also shown in the SMM plots. Vertical lines indicate the classical $2$ and $10\,U_\text{p}$ cutoffs. (g-i) Total electron yields as function of relative phase extracted from TDSE (gray areas) and SMM (black curves) and selective yields from TDSE for the energy regions as indicated and visualized at the top of panel (a). Note that the plots are normalized to the respective peak values.}
\label{fig:tdse_smm_spec}
\end{figure}

Full phase-dependent ATI spectra for two-color excitation as predicted from the quantum and classical models are shown in figure~\ref{fig:tdse_smm_spec} for representative values of the decay length. In all cases, the quantum mechanical model predicts strong modulations of the ATI peaks with relative phase, see figure~\ref{fig:tdse_smm_spec}(a)-(c). By fitting a harmonic oscillation $Y_{\mathrm{fit}}=A+B\cos(\varphi-\varphi_{\mathrm{crit}})$ to the phase-dependent yield of individual ATI peaks, we determine energy-dependent critical phases $\varphi_\text{crit}(E)$ that characterize maximal two-color yield (black curves in figure~\ref{fig:tdse_smm_spec}(a)-(f)). Note that the resulting evolution of critical phases exhibits a chirp from values of $1.2\pi$ in the low energy region to $0.8\pi$ in the plateau region. Besides the modulation of individual ATI peaks in the low and intermediate energy range, also the high energy cutoff of the ATI spectra itself becomes modulated and results in a phase evolution of the yield with a critical phase (maximal cutoff) around $\varphi_\text{crit}=1.5\pi$. In addition, while the overall structure persists, the cutoff is gradually shifted to lower energies as the near-field becomes more localized. It should be emphasized that the data from the quantum model alone does not allow a rigorous and intuitive further identification of the underlying mechanisms.\\

\begin{figure}[h]
\centering
\includegraphics[width=0.5\textwidth]{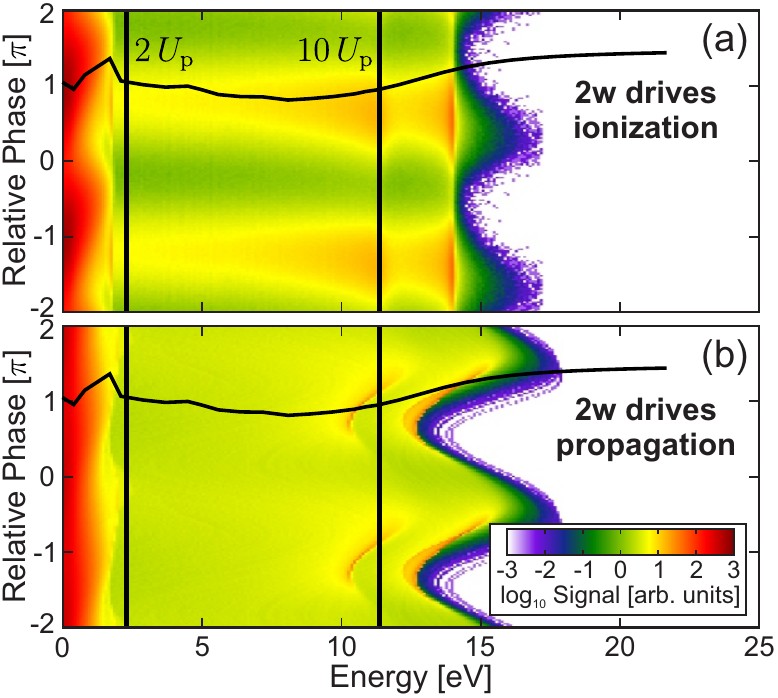}
\caption{Energy spectra predicted by SMM as function of relative phase for selectively activated $2\omega$-fields for decay length $\lambda_\text{nf} = \unit[100]{\AA}$. In the upper panel (a) the $2\omega$-field is included in the evaluation of the ionization rate but neglected in the trajectory integration. In panel (b) only the fundamental field is used to evaluate ionization but the full two-color field is used to integrate the trajectories. Black curves show the critical phases as function of energy as extracted from TDSE results (cf. figure~\ref{fig:tdse_smm_spec}b). }
\label{fig:tdse_smm_spec_selective}
\end{figure}

The corresponding SMM results in figure~\ref{fig:tdse_smm_spec}(d)-(f) show a clear direct electron feature, a relatively smooth plateau with step features associated to individual half-cycles near the pulse peak, and a sharp global cutoff. Although the formation of ATI peaks cannot be explained classically, several trends of the quantum result are well captured by the SMM model: (i) the shape and phase evolution of the direct electron feature agrees very well, (ii) in the plateau region, maximal yield is found for phases similar to the respective critical phases of the quantum version, (iii) the modulation near the cutoff is qualitatively captured. As the ionization step in the SMM can be disentangled from the trajectory integration, a selective activation of the $2\omega$-field can be performed and uncovers the impacts of the trajectory effects and the ionization rate effect on the two-color spectra.

A selective analysis for the case with moderate decay length from figure~\ref{fig:tdse_smm_spec} is given in figure~\ref{fig:tdse_smm_spec_selective}. This analysis shows that the modulation signatures at low and intermediate energies are essentially captured completely by activating the $2\omega$ field only in the ionization rate (figure~\ref{fig:tdse_smm_spec_selective}(a)) and are practically absent when activating the $2\omega$ field only in the trajectory integration (figure~\ref{fig:tdse_smm_spec_selective}(b)). Hence, it can be concluded that the trajectory effect is negligible for the two-color effect in this spectral region. In contrast to that, capturing the high energy cutoff modulation is possible only when using the full two-color field in the trajectory integration but not necessarily in the ionization. Note that also the ionization rate effect results in a small cutoff modulation due to mere statistical weight of corresponding trajectories. This cutoff modulation, however, exhibits the wrong phase behavior when compared to TDSE. Therefore our analysis clearly supports that the phase-sensitivity of the two-color photoelectron spectra results from and can be controlled via fundamentally different effects. Whereas the ionization effect governs the $2\omega$-contrast in low and intermediate energies, the cutoff region is most sensitive to the two-color effect on the trajectory. The fact that the critical phases in the cutoff region are robust against changes of the decay length shows that the two-color modulation of the signal in this region may be a suitable marker for the characterization of the relative phase of two-color fields.

\section{Conclusion}
We have analyzed the two-color photoemission from sharp nanotips within a simplified one-dimensional model and have performed simulations for parameters motivated by a particular experimental scenario. We find clear two-color signatures and identify a quenching of the quiver motion as well as a ponderomotive energy shift effect as a result of the finite near-field extension. Most importantly, our analysis of the phase-dependent yield within the quantum and classical versions of the model shows that the plateau resulting from elastic backscattering obtains its phase sensitivity because of two different effects, namely the ionization rate modification and the trajectory modification. In particular the cutoff modulation is captured well classically and proofs to be very robust, supporting its relevance for the measurement and calibration of the relative two-color phase in a nanotip phasemeter device.

\ack
We gratefully acknowledge financial support from the Deutsche Forschungsgemeinschaft within SPP 1840 and via a Heisenberg fellowship (T.F.). Computer time was provided by the North-German Supercomputing Alliance HLRN (project ID mvp00011). T.P. and P.H. further acknowledge funding through ERC Consolidator Grant NearFieldAtto.

\providecommand{\newblock}{}

\end{document}